\documentclass[twocolumn,showpacs,preprintnumbers,amsmath,amssymb,prl]{revtex4-1}

\usepackage{graphicx}
\usepackage{dcolumn}
\usepackage{bm}
\usepackage{graphics}
\usepackage{subfigure}
\usepackage{graphicx}
\usepackage{epsfig}
\usepackage{epstopdf}
\usepackage{color}
\usepackage{xcolor}

\begin{document}

\title{Why does graphene behave as a weakly interacting system?}

\author{Johannes Hofmann}

\email{hofmann@umd.edu}

\author{Edwin Barnes}

\author{S. Das Sarma}

\affiliation{Condensed Matter Theory Center and Joint Quantum Institute, Department of Physics, University of Maryland, College Park, Maryland 20742-4111 USA}

\date{\today}

\begin{abstract}
We address the puzzling weak-coupling perturbative behavior of graphene interaction effects as manifested experimentally, in spite of the effective fine structure constant being large, by calculating the effect of Coulomb interactions on the quasiparticle properties to next-to-leading order in the random phase approximation (RPA). The focus of our work is graphene suspended in vacuum, where electron-electron interactions are strong and the system is manifestly in a nonperturbative regime. We report results for the quasiparticle residue and the Fermi velocity renormalization at low carrier density. The smallness of the next-to-leading order corrections that we obtain demonstrates that the RPA theory converges rapidly and thus, in contrast to the usual perturbative expansion in the bare coupling constant, constitutes a quantitatively predictive theory of graphene many-body physics for any coupling strength.
\end{abstract}

\maketitle

Graphene, a single-atom thick sheet of graphite, consists of carbon atoms arranged in a two-dimensional honeycomb lattice structure. Valence and conduction bands touch only at two Dirac points at the corners of the Brillouin zone, giving rise to a linear relativistic dispersion relation of the low-energy excitations $\varepsilon({\bf q}) = v_F |{\bf q}|$, where the speed of light $c$ is replaced by the Fermi velocity $v_F \approx c/300$~\cite{wallace47}. Hence, graphene displays many effects found in relativistic field theories, such as the Klein paradox~\cite{katsnelson06} or the half-integer quantum Hall effect~\cite{novoselov05}. Because of these novel electronic properties and possible technological applications, graphene has developed into one of the most active areas of physics research over the past decade~\cite{novoselov04,novoselov05,dassarma11,kotov12}.

The electrons in graphene have a mutual Coulomb interaction with a strength characterized by the dimensionless ratio of potential to kinetic energy (i.e., the effective graphene fine structure constant):
\begin{align}
\alpha &= \frac{e^2}{\kappa v_F} , \label{eq:alpha}
\end{align}
where $\kappa$ is the background dielectric constant arising from the surrounding medium. We set $\hbar=1$ throughout this Letter. Since $v_F \ll c$, retardation effects are negligible and the Coulomb interaction is instantaneous, thus breaking the emergent relativistic invariance of the noninteracting theory. For graphene suspended in vacuum, the interaction strength is $\alpha=2.2$. Clearly, this is not small, and \emph{a priori}, there is no reason why a perturbative expansion in $\alpha$ should give reliable results when computing physical quantities. (By contrast, $\alpha$=1/137 in quantum electrodynamics and therefore the weak-coupling perturbation theory in $\alpha$ remains valid up to many orders.) If the chemical potential is tuned away from half filling (extrinsic graphene), the strong long-ranged interaction is screened and the system behaves as a weakly interacting Landau Fermi liquid as in ordinary 2D and 3D clean metals~\cite{shankar94}. However, as the density is decreased towards the Dirac point (intrinsic graphene), the interaction cannot be screened any longer. Even though a quasiparticle description persists, there are strong renormalization effects on the Fermi velocity $v_F$ and the quasiparticle residue $Z$~\cite{kotov12}.

The quasiparticle renormalization has been determined through measurements of the effective cyclotron mass~\cite{elias11}, the infrared conductivity~\cite{li08}, scanning tunneling spectroscopy~\cite{li09,chae12} and also through a direct angle-resolved photoemission spectroscopy measurement of the Dirac cones~\cite{siegel11}. All experiments detect an increase in the Fermi velocity close to the Dirac point. The strongest renormalization is reported in the experiment by Elias \emph{et al.}~\cite{elias11}, who observe a Fermi velocity enhancement by a factor of $3$ at low carrier density. Theoretically, the scale dependence of the Fermi velocity arises from the renormalization of the graphene effective field theory, where the unphysical dependence of bare correlation functions on an ultraviolet regulator induces a scale-dependent renormalized Fermi velocity. The first perturbative calculation of the quasiparticle renormalization dates back to the work of Gonz\'alez \emph{et al.}~\cite{gonzalez94,gonzalez99,gonzalez01}, which predicts a logarithmic enhancement at small density within a leading-order (LO) perturbative calculation in the bare coupling.

So far, all existing experimental measurements report a Fermi velocity renormalization that is consistent with simple first-order perturbation theory~\cite{gonzalez94} despite the fact that the interaction strength is not small in real graphene. Notwithstanding extensive works on this subject, this apparent consistency has remained an outstanding theoretical puzzle. One possibility is that the agreement between theory and experiment is simply fortuitous: if the perturbative renormalization group is carried out to second order, an unphysical fixed point in the Fermi velocity beta function appears at a critical coupling of $\alpha^* = 0.78$~\cite{barnes14}. For suspended graphene, this indicates a decrease in the Fermi velocity at low density, which is in disagreement with the experimental findings. In fact, Dyson's original argument for the breakdown of perturbation theory applied to graphene indicates that this may happen already at first or second order in $\alpha$, thus implying the failure of perturbation theory for graphene in vacuum~\cite{barnes14}. These results show that perturbation theory does not provide a quantitatively (or perhaps even qualitatively) reliable theory of graphene in vacuum or on most substrates. We therefore have a conundrum where theory indicates an explicit failure of perturbation theory whereas experiments report agreement with the simplest leading-order perturbative result.

In this Letter, we demonstrate explicitly that the random phase approximation (RPA) yields a quantitatively predictive theory of graphene many-body effects even in the case of strong bare interactions (i.e., $\alpha >0.78$) such as in suspended graphene (or graphene on SiO${}_2$ substrates, the most-studied graphene system in the literature, where $\alpha=0.9$). We establish this by going to next-to-leading order (NLO) in the RPA expansion, where we find that the corrections to the Fermi velocity and quasiparticle residue are small relative to the leading-order RPA results. This shows that the RPA expansion is not only systematic and well behaved, but that it is also quickly convergent, in sharp contrast to ordinary perturbation theory. In addition to supplying a proper theoretical justification for the RPA theory, the subleading corrections obtained in this work allow for a more detailed, quantitative comparison between theory and future experiments.

\begin{figure}
\subfigure[]{\scalebox{0.5}{\includegraphics{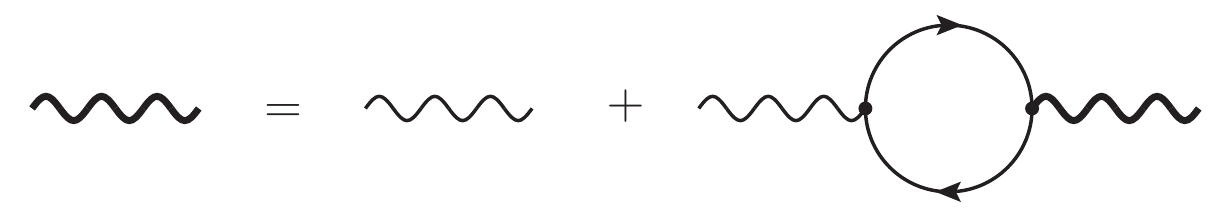}}\label{fig:RPA}}\\
\subfigure[]{\scalebox{0.5}{\includegraphics{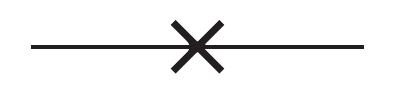}}\label{fig:ct1}}
\subfigure[]{\scalebox{0.5}{\includegraphics{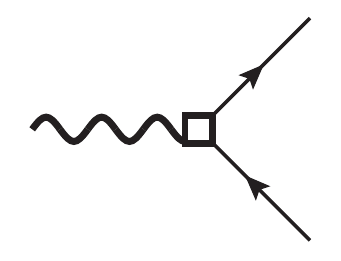}}\label{fig:ct2}}\\
\subfigure[]{\scalebox{0.5}{\includegraphics{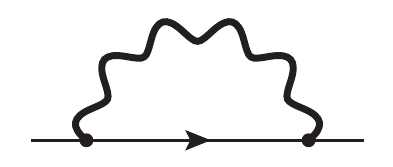}}\label{fig:se1}}
\subfigure[]{\scalebox{0.5}{\includegraphics{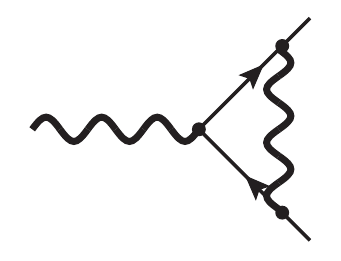}}\label{fig:v1}}\\
\subfigure[]{\scalebox{0.5}{\includegraphics{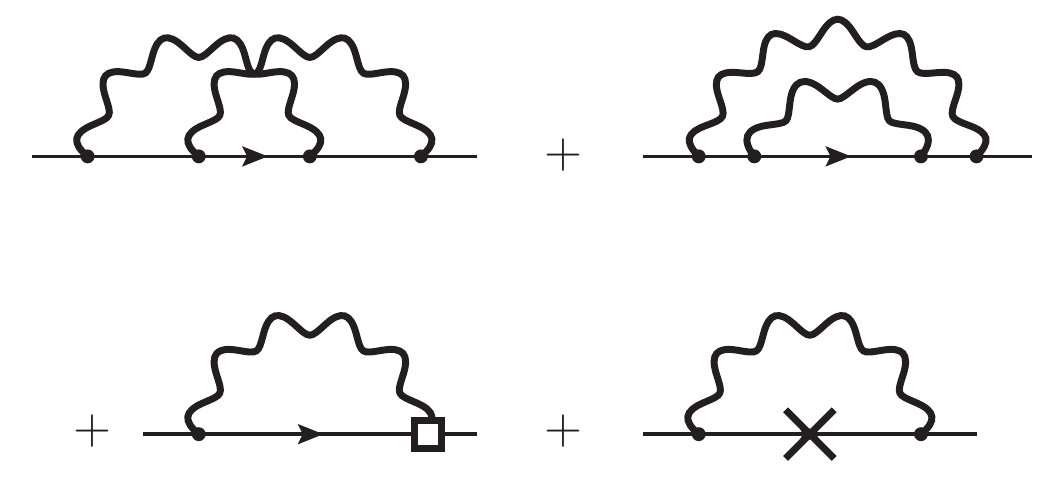}}\label{fig:NLO}}
\caption
{
(a) Schwinger-Dyson equation for the RPA interaction. (b),(c) Diagrams that renormalize the propagator (b) and the vertex (c) to leading order in the RPA. (d),(e) The logarithmic divergences in perturbation theory are subtracted by propagator and vertex counterterms, respectively, as defined in Eq.~\eqref{eq:counter}. (f) Diagrams that renormalize the wave function and the Fermi velocity to next-to-leading order in the RPA.
}
\label{fig:fey}
\end{figure}

The low-energy behavior of graphene can be described by the following action in Euclidean space:
\begin{align}
S &= - \int dt \int d^2x  \left(\bar{\psi}_a \gamma^0 \partial_t \psi_a {+} v_F \bar{\psi}_a \gamma^i \partial_i \psi_a {+} A_0 \bar{\psi}_a \gamma^0 \psi_a\right) \nonumber \\
&+ \frac{1}{2g^2} \int dt \int d^2x \int dz \, (\partial_i A_0)^2 , \label{eq:action}
\end{align}
where the fermion spinors $\psi_a$ are four-component fields corresponding to the sublattice and valley degrees of freedom and we leave a summation over the spin multiplicity $a=1,\ldots,N$ implicit. The Dirac $\gamma$ matrices obey the Clifford algebra $\{\gamma^\mu, \gamma^\nu\} = 2 \delta_{\mu\nu}$, and we define $\bar{\psi}_a = \psi_a^\dagger \gamma^0$. The non dynamical gauge field $A_0$ mediates the instantaneous three-dimensional Coulomb interaction. The coupling $g^2$ is related to the graphene fine-structure constant by $\alpha = g^2/4\pi v_F$. The RPA interaction is obtained by summing a geometric series of Coulomb lines and polarization loops as shown in Fig.~\ref{fig:RPA}, which gives
\begin{align}
V_{\rm RPA}(q) &= 2 \pi \alpha {v_F} \biggl(|{\bf q}| + \alpha{v_F} \frac{\pi N}{4} \frac{|{\bf q}|^2}{\sqrt{q^2}}\biggr)^{-1} , \label{eq:rpa}
\end{align}
where ${\bf q}$ denotes the momentum component of the Euclidean three-vector $q$, and $q^2 = q_0^2 + {v_F^2}{\bf q}^2$. We emphasize that in our nomenclature RPA implies an expansion in the screened Coulomb interaction $V_{\rm RPA}$ [given in Eq.~\eqref{eq:rpa} and in the bold wavy line in Fig.~\ref{fig:fey}] whereas perturbation theory implies an expansion in the bare Coulomb interaction [given by the first term on the right hand side of Eq.~\eqref{eq:rpa} and the nonbold wavy line in Fig.~\ref{fig:RPA}].  In the current work, graphene many-body renormalization is explicitly calculated to the second order in $V_{\rm RPA}$ whereas Barnes \emph{et al.}~\cite{barnes14} carried out the bare perturbation theory to second order.

Because of the slow $1/q$ decay of the Coulomb interaction in momentum space, correlation functions computed using the action~\eqref{eq:action} are sensitive to high-momentum modes and display logarithmic divergences. We regulate the theory~\eqref{eq:action} using a hard cutoff $\Lambda$ in momentum space and then subtract the logarithmic divergences by adding a counterterm action
\begin{align}
S_{\rm c.t.} &= \sum_{a=1}^N \int d^dx \, \bigl(A \, \bar{\psi}_a \gamma^0 \partial_t \psi_a + v_F B \, \bar{\psi}_a \gamma^i \partial_i \psi_a \nonumber \\
&\qquad\qquad + C A_0 \bar{\psi}_a \gamma^0 \psi_a\bigr) \label{eq:counter}
\end{align}
to Eq.~\eqref{eq:action}. Diagrammatically, the counterterms induce additional two-particle and vertex functions shown in Figs.~\ref{fig:ct1} and~\ref{fig:ct2}. Correlation functions computed including the counterterms are manifestly free of divergences. Since the electron charge $e^2$ is not renormalized~\cite{ye98,herbut06}, any divergence in the electron-Coulomb vertex is removed by a simple wave function renormalization, which implies $A=C$. The wave function and Fermi velocity renormalization can be determined from a computation of the self-energy. In a renormalization scheme in which only the divergent pieces are subtracted, the counterterm coefficients can be expanded as
\begin{align}
A(\alpha) &= \sum_{n=1}^\infty a_n(\alpha) \log^n \frac{\Lambda}{\mu} \label{eq:A} \\
B(\alpha) &= \sum_{n=1}^\infty b_n(\alpha) \log^n \frac{\Lambda}{\mu} , \label{eq:B}
\end{align}
where $\mu$ is an arbitrary renormalization scale with dimension of momentum. The running of the scale-dependent renormalized Fermi velocity $v_F(\mu)$ and field strength $Z(\mu)$ is governed by the leading-order term in Eqs.~\eqref{eq:A} and~\eqref{eq:B}~\cite{amit05}:
\begin{align}
\mu \frac{dZ}{d\mu} &= 2 \gamma_\psi(\alpha) Z = a_1(\alpha) Z  \\
\mu \frac{dv_F}{d\mu} &= \gamma_F(\alpha) v_F = [a_1(\alpha) - b_1(\alpha)] v_F. \label{eq:flow}
\end{align}
These equations allow us to relate the Fermi velocity and field strength at different momentum scales. It turns out that the coefficient on the right-hand side of Eq.~\eqref{eq:flow} is negative, and hence the Fermi velocity grows as the momentum scale decreases. This implies that the dimensionless Coulomb interaction strength~\eqref{eq:alpha} vanishes at small scales; i.e., the Coulomb interaction is marginally irrelevant in a renormalization group (RG) sense. In practice, the renormalization group flow is stopped at a characteristic momentum scale, which at low temperature can be identified with the Fermi momentum $k_F = \sqrt{\pi n}$.

\begin{figure}[t]
\scalebox{0.65}{\includegraphics{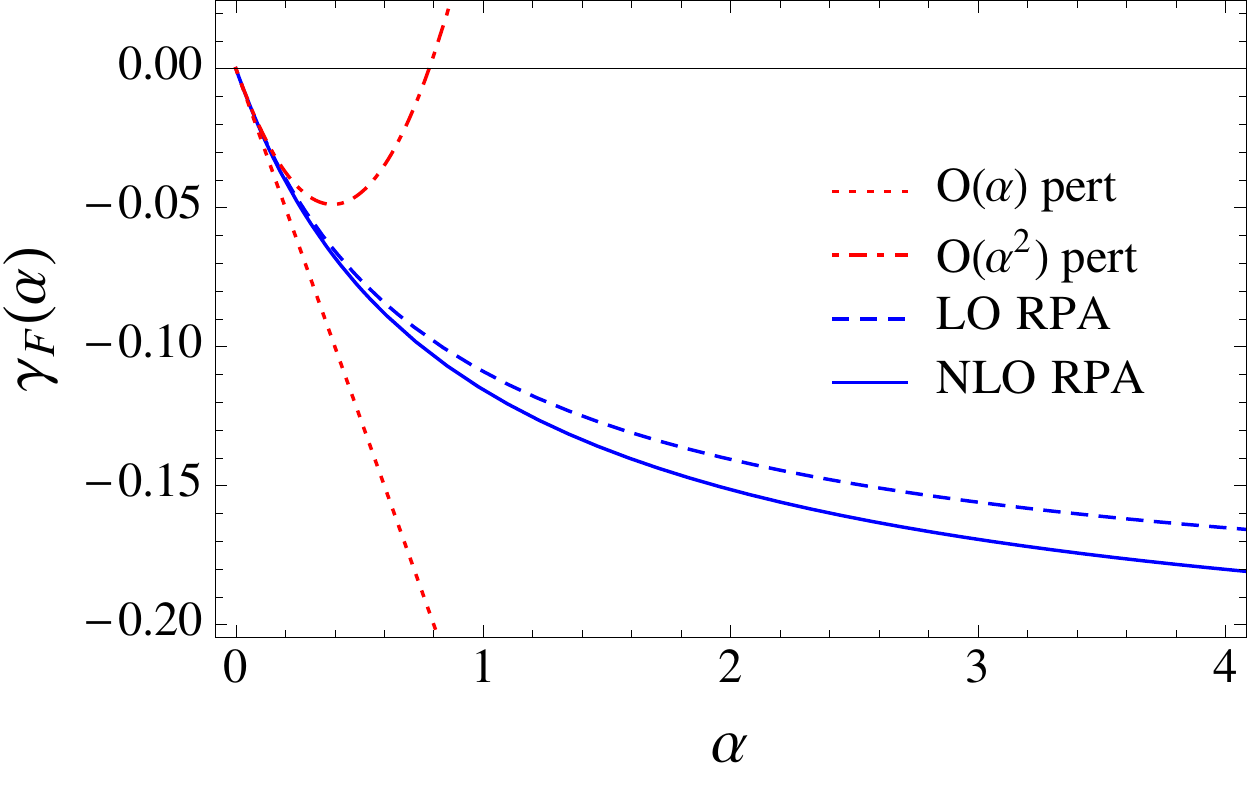}}
\caption
{
Anomalous dimension of the Fermi velocity as a function of the fine structure constant $\alpha$. The dashed blue line is the leading-order RPA result from Ref.~\cite{son07}. The solid blue line is the result including the next order in the RPA as obtained in this Letter. The correction is small and changes the quantitative features of the RG flow. In contrast, perturbation theory at second order in $\alpha$ (red dashed dotted) gives a strikingly different result from the first-order (red dashed) flow, and predicts an unphysical fixed point at $\alpha^* = 0.78$~\cite{barnes14}.
}
\label{fig:rg}
\end{figure}

The Feynman diagrams of the self-energy and the vertex to leading order in the RPA interaction are depicted in Figs.~\ref{fig:se1} and~\ref{fig:v1}. They define the counterterms $a_1(\alpha)$ and $b_1(\alpha)$ to leading order, for which the explicit results can be evaluated analytically and will not be repeated here~\cite{dassarma07,son07}. The main point of this work is to go beyond this leading-order calculation and to compute the running of the residue and Fermi velocity to next-to-leading order in the RPA. There are four diagrams that contribute to the renormalization group flow, which are shown in Fig.~\ref{fig:NLO}. The first two diagrams include vertex and self-energy insertions in the leading-order RPA diagram. In addition, the counterterms as computed to leading order also renormalize the subdivergences that arise at next-to-leading order. We decompose the self-energy as $\Sigma(q) = \Sigma_0 \gamma^0 q_0 + \Sigma_1 {\bf q} \cdot \boldsymbol{\gamma}$ and project each diagram on the scalar components $\Sigma_0$ and $\Sigma_1$. After imposing a cutoff, the resulting finite expression is computed using an exact quadrature rule for one-loop expressions and the Vegas algorithm for two-loop diagrams as implemented in the CUBA library~\cite{hahn05}. Varying the cutoff over several orders of magnitude, we can extract the logarithmic divergences. Note that at next-to-leading order, each diagram has double-logarithmic divergences which can be computed in closed analytical form starting with the known results for the one-loop self-energy and vertex correction. Our numerical results agree with these analytical findings, thus providing an independent check of our computation. Figure~\ref{fig:rg} shows the result for the Fermi velocity anomalous dimension, where it is apparent that while first- and second-order perturbation theories give vastly different results, the NLO RPA calculation provides a small correction to the LO RPA results. Thus, the RPA expansion in $V_{\rm RPA}$ converges well in contrast to perturbation theory in the bare coupling $\alpha$.

We now compare our result from the NLO RPA with existing experimental measurements and contrast it with the ordinary perturbation theory results. The renormalized Fermi velocity at a certain momentum scale can be obtained by integrating the renormalization group equation~\eqref{eq:flow} starting with an initial value of the Fermi velocity at high density. Note that we neglect an additional correction that is small compared to the dominant electron-electron renormalization effect, and which arises from the fact that we expand around the Dirac point and not the Fermi surface~\cite{dassarma07}.  
\begin{figure}[t]
\scalebox{0.65}{\includegraphics{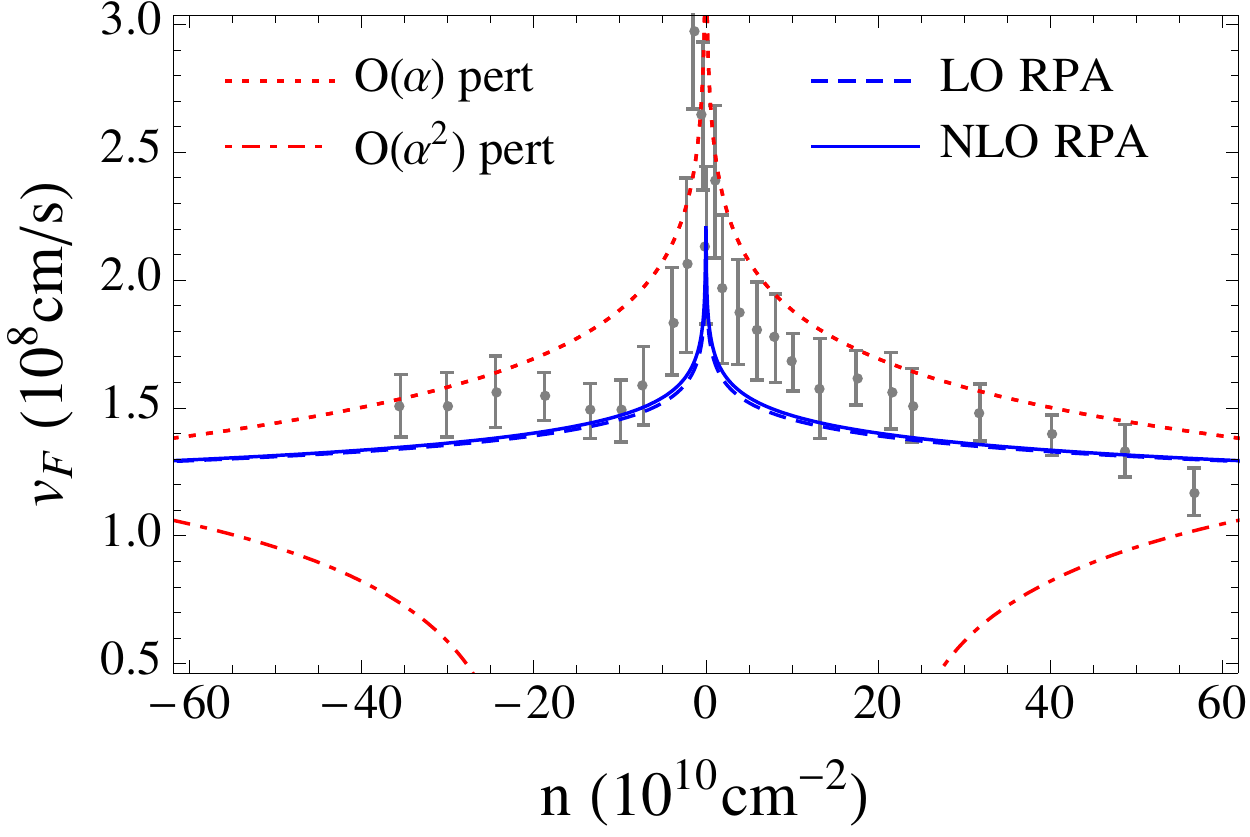}}
\caption
{
Renormalization of the Fermi velocity in suspended graphene. The Fermi velocity is obtained by integrating the flow equations~\eqref{eq:flow} starting at a density $n_0 = 100 \times 10^{10} {\rm cm}^{-2}$ with $v_F^0 = 1.24 \times 10^8 {\rm cm}/{\rm s}$. While first-order perturbation theory (red dashed) is qualitatively consistent, the second-order result (red dashed dotted) is in striking disagreement with the experimental data. In contrast, the LO (blue dashed) and NLO (blue solid) RPA results are both in close agreement. The experimental data points are taken from Ref.~\cite{elias11}.
}
\label{fig:renvf}
\end{figure}
Figure~\ref{fig:renvf} presents our results for the Fermi velocity renormalization in suspended graphene. The Fermi velocity beta function is solved with the initial condition $v_F^0 = 1.24 \times 10^{8}{\rm cm/s}$ at $n_0 = 10^{12} \, {\rm cm}^{-2}$ as quoted in Ref.~\cite{elias11}. First, consider the perturbative results obtained in Ref.~\cite{barnes14}. As discussed in the introduction, the first-order result shown in Fig.~\ref{fig:renvf} is in agreement with the experimental data obtained in Ref.~\cite{elias11}. The second-order contributions, however, lead to results that are in complete disagreement with the experiment. In contrast, the RPA calculation does not exhibit such ambiguities: as is clear from the figure, the NLO calculation gives a small correction to the leading-order renormalization. The agreement of the NLO velocity with both the experimental data and the LO result provides strong evidence that the RPA is indeed a well-defined systematic expansion that describes the properties of suspended graphene both quantitatively and qualitatively. This is the main result of the present work.

\begin{figure}[t!]
\scalebox{0.65}{\includegraphics{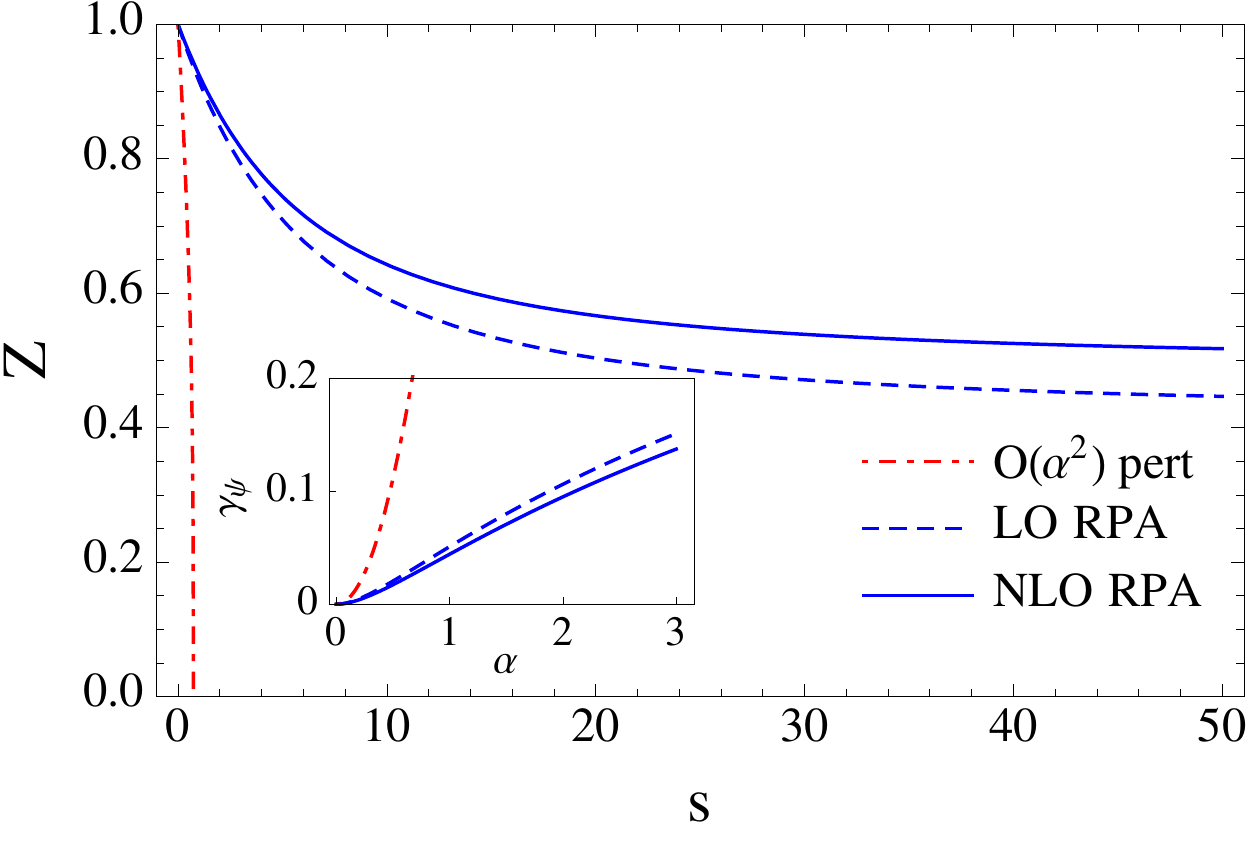}}
\caption
{
Running of the quasiparticle residue for suspended graphene as the initial scale $\mu_0 \sim \sqrt{n_0}$ is reduced to $\mu(s) = e^{-s} \mu_0$. The residue is not renormalized to zero but approaches a constant value in the low-density limit. The initial conditions for the flows are the same as in Fig.~\ref{fig:renvf}.
}
\label{fig:Zren}
\end{figure}

The fermionic quasiparticles in intrinsic graphene display a strange Fermi liquid behavior: while conventional Fermi liquid theory predicts an inverse lifetime $1/\tau \sim \omega^\nu$ with $\nu > 1$, it was shown that the decay rate is linear in energy, very much like for a marginal Fermi liquid~\cite{gonzalez99,dassarma07}. The quasiparticle residue, on the other hand, does not renormalize to zero as shown in Fig.~\ref{fig:Zren}, which displays the integrated RG flow of the residue for suspended graphene obtained with the same initial conditions as in Fig.~\ref{fig:renvf}. Rather, we see that the leading-order RPA predicts a finite $Z$ even if the system is tuned exactly to the Dirac point, as noted previously in Ref.~\cite{gonzalez99}. As before, including the NLO RPA correction induces only a small quantitative change, as is clear from the figure. We also include the perturbative result in Fig.~\ref{fig:Zren} for comparison. To leading order in $\alpha$, there is no wave function renormalization, whereas at second order, the quasiparticle residue quickly goes to zero and acquires unphysical negative values, which reflects the breakdown of perturbation theory.

Finally, we briefly comment on the strong-coupling behavior of the graphene effective theory. To leading order in a large-$N$ treatment, which coincides with the LO RPA~\cite{vafek07,son07,drut08,foster08}, the renormalization group flow possesses an infrared repulsive fixed point at infinite Coulomb interaction~\cite{son07}. It has been argued that the anomalous dimension at this strong-coupling fixed point has potentially observable consequences~\cite{son07,drut08}, in particular, dictating a fermion dispersion relation at large momenta $\sim p^z$ with an anomalous dimension $z<1$. Our calculation contributes to the NLO ${\cal O}(1/N^2)$ in the large-$N$ expansion, and we quote the numerical result of this correction:
\begin{align}
z = 1 - \frac{4}{\pi^2 N} - \frac{0.85(20)}{(\pi N)^2} + {\cal O}(1/N^3) ,
\end{align}
where the leading order ${\cal O}(1/N)$ was obtained in~\cite{son07}. For $N=2$, corresponding to the physical case, we find a small correction of the order of $2\%$. Note that the NLO receives an additional correction that corresponds to a higher order three-loop RPA diagram with a vertex or self-energy insertion in the polarization bubble. We anticipate this correction to be of the same magnitude or less compared with the two-loop RPA diagrams we have calculated here. Our result for the anomalous dimension $z$ suggests that the true expansion parameter is $1/\pi N$ rather than $1/N$, which implies that the large-$N$ expansion is reliable even for strong interactions and a seemingly small spin degeneracy of $N=2$. This effective hidden expansion parameter of $1/2\pi$ may be the qualitative reason for the RPA expansion to be convergent even for manifestly strong coupling $\alpha=2.2$.

In conclusion, motivated by the fact that perturbation theory is inapplicable to suspended graphene, we systematically compute the second-order correction to the renormalization group flow using the RPA interaction. Corrections to the leading-order result are small and our results are in good agreement with recent measurements of the Fermi velocity renormalization in suspended graphene. Our investigation provides conclusive evidence that the random phase approximation gives a quantitatively and qualitatively accurate description of graphene even at strong coupling. Our results constitute quantitative predictions for future, higher-precision experiments measuring graphene many-body renormalization effects.

This work is supported by JQI-NSF-PFC and ARO-MURI.

\bibliography{bib}

\end{document}